\documentclass[conference]{IEEEtran}
\IEEEoverridecommandlockouts
\usepackage{cite}
\usepackage{amsmath,amssymb,amsfonts}
\usepackage{algorithmic}
\usepackage{graphicx}
\usepackage{textcomp}
\usepackage{xcolor}
\def\BibTeX{{\rm B\kern-.05em{\sc i\kern-.025em b}\kern-.08em
    T\kern-.1667em\lower.7ex\hbox{E}\kern-.125emX}}
\begin{document}

\title{Toward Deep Learning-based Segmentation and Quantitative Analysis of Cervical Spinal Cord Magnetic Resonance Images\\
}
\author{\IEEEauthorblockN{Maryam Tavakol Elahi}
\IEEEauthorblockA{\textit{School of Electrical Engineering and Computer Science} \\
\textit{University of Ottawa}\\
Ottawa, Canada \\
mtava020@uottawa.ca}
}

\maketitle

\begin{abstract}
  This research proposal discusses two challenges in the field of medical image analysis: the multi-parametric investigation on microstructural and macrostructural characteristics of the cervical spinal cord and deep learning-based medical image segmentation. First, we conduct a thorough analysis of the cervical spinal cord within a healthy population. Unlike most previous studies, which required medical professionals to perform functional examinations using metrics like the modified Japanese Orthopaedic Association (mJOA) score or the American Spinal Injury Association (ASIA) impairment scale, this research focuses solely on Magnetic Resonance (MR) images of the cervical spinal cord. Second, we employ cutting-edge deep learning-based segmentation methods to achieve highly accurate macrostructural measurements from MR images. To this end, we propose an enhanced UNet-like Transformer-based framework with attentive skip connections. 
  This paper reports on the problem domain, proposed solutions, current status of research, and expected contributions.
  
\end{abstract}

\begin{IEEEkeywords}
Deep Learning, Medical Image Segmentation, Transformer, Skip Connection, Cervical Spinal Cord, Magnetic Resonance Imaging, Diffusion Tensor Imaging, Fractional Anisotropy, Quantitative Analysis, Cervical Stenosis
\end{IEEEkeywords}

\section{Introduction and Motivation}
Medical images are highly expressive and useful, but the critical details and visual characteristics they encompass are rarely discernible to even the most trained eyes. Knowing that various types of medical images, especially MRI and CT scans, are among the essential and fundamental resources for acquiring relevant information about patients' bodies and different diseases, interpreting them remains a top priority. 

As a small and crucial part of the central nervous system, the spinal cord has greatly benefited from MRI's non-invasive and superior soft tissue contrast capabilities. Over the past two decades, MRI has progressed from structural to quantitative imaging, allowing for both qualitative and quantitative assessments of spinal cord tissue structure. 

Technological innovations, such as diffusion tensor imaging (DTI) and the development of high-field MRI scanners, have moved quantitative imaging forward, enabling more detailed examinations of tissue microstructure. Building on this progress, now is the time to delve deeper into the spinal cord's microstructure using advanced quantitative MR imaging techniques.
This approach not only assists clinicians in tracking disease progression and predicting clinical outcomes but also facilitates earlier disease diagnosis by identifying quantitative neurological changes that precede structural alterations. Recent literature highlights the potential of DTI metrics as valuable biomarkers for early disease detection. In light of these advancements, our study focuses on utilizing quantitative measurements to evaluate the microstructural and macrostructural characteristics of the cervical spinal cord MR images. 

Microstructural features refer to the fine details of spinal cord tissue, which can be analyzed at the microscopic level. They encompass elements such as nerve fibers, axons, and the myelin sheath surrounding the axons. Fractional anisotropy (FA), for instance, is a microstructural characteristic that can provide insights into fiber density, axonal diameter, and myelination in white matter. Variations in these microstructural elements can indicate damage or disease, often manifested in symptoms of pain, numbness, or motor function loss. On the other hand, macrostructural features focus on the more observable aspects of the spinal cord, including size, shape, volume, and other anatomical details. Macrostructural changes can be indicative of larger-scale problems such as spinal cord injury, deformity, or compression.

In this study, we aim to contribute to the automatic analysis and interpretation of medical images with the spinal cord as an exemplar by developing applicable methodologies. 
We seek to investigate the following research questions

\noindent\textbf{RQ1:} 
How can the analysis and interpretation of medical images, specifically CT and MRI scans, be improved and automated using state-of-the-art deep learning-based computer vision methods?
    
\noindent\textbf{RQ2:} 
How can we find relationships between microstructural and macrostructural features of the cervical spinal cord in a healthy population by eliminating the need for subjective measurements from clinicians, thereby objectively understanding these relationships using the acquired MR images purely?
    
\noindent\textbf{RQ3:}
How can we achieve highly accurate macrostructural measurements from cervical spinal cord MR images by leveraging the capabilities of recent deep learning-based segmentation methods?

This research is conducted in three steps. First, addressing \textbf{RQ1}, understanding the cervical spinal cord's microstructural and macrostructural characteristics and their interrelations, we aim to improve and automate the extraction and analysis of these features from MR images, using the spinal cord as an exemplar.

Second, for \textbf{RQ2}, we present a multiparametric approach for evaluating the microstructural and macrostructural features of the cervical spinal cord in healthy individuals. We delve into the investigation of the correlation between them while concurrently examining the influence of gender and different imaging machines on these correlations.
The proposed approach aims to establish relationships between these features, thereby fostering a deeper understanding of the microstructural and macrostructural characteristics and examining whether the microstructural changes might occur from various degrees of asymptomatic stenosis. 
All these measurements are captured using quantitative MRI, eliminating the need for subjective assessments and relying primarily on objective MRI data.

In the third step, addressing \textbf{RQ3}, we propose an enhanced UNet-like Transformer-based framework with attentive skip connections for high-performance image segmentation. Our approach features a novel Transformer-based skip connection module that integrates features extracted from both the encoder and decoder, enabling it to capture more complex dependencies between different levels of abstraction. 
We further improve the framework's efficiency and its ability to process high-resolution images by adopting a merging cross-covariance attention mechanism in place of the conventional self-attention operation.

The rest of this paper is organized as follows. Section~\ref{sec:relatedWork} discusses related work, and Section~\ref{sec:solutions} presents the proposed multi-parametric method to evaluate the cervical spinal cord with a focus on quantitative measurements and the proposed deep learning-based segmentation framework. Expected contributions are outlined in section~\ref{sec:contributions}, and section~\ref{sec:conclusion} concludes and highlights future work.

\section{Related Work}
\label{sec:relatedWork}
\subsection{Cervical spinal cord disease analysis}
In a study, Avinash et al.~\cite{rao2018diffusion} investigate the usefulness of FA as a biomarker for the severity of cervical spondylotic myelopathy (CSM) patient cases and as a prognostic biomarker for improvement after surgery. The regression analysis performed between FA and mJOA score indicates that FA at the level of maximal compression (LMC) correlates positively with pre-operative mJOA score, and pre-operative FA correlates inversely with recovery throughout the post-operative period.

In another study~\cite{kitamura2020longitudinal}, the analysis of correlations between the pre- and post-operative FA and mean diffusivity (MD) values and the pre- and post-operative JOA scores reveals that although the JOA score improved significantly after surgery, no significant changes were observed in the pre- and post-operative FA and MD values. 

Another work~\cite{ellingson2018reproducibility} quantifies the reproducibility, temporal stability, and functional correlation of diffusion MR characteristics in the spinal cord of patients with cervical stenosis with or without myelopathy. The study explores the association between longitudinal DTI measurements and serial neurological function assessment. 
Their research concerning the specific DTI measurements shows that FA within the spinal cord appears slightly more sensitive to neurological function and more stable than measures of MD. 

Kara et al.~\cite{kara2011role} show that DTI may detect abnormalities in the spinal cord before the development of T2 hyper-intensity on conventional sequences in patients with CSM. 

Other studies have examined the efficacy of diffusion tensor imaging in recognizing degenerative CSM. For instance, Orel et al.~\cite{zaninovich2019role} investigates the role of diffusion tensor imaging in the diagnosis, prognosis, and assessment of recovery and treatment of spinal cord injury. It is demonstrated that DTI metrics and combinations thereof correlate significantly with clinical function in both model species and humans.

\subsection{Deep learning-based medical image analysis}

U-Net~\cite{ronneberger2015u} is a widely-used CNN-based medical image segmentation method that utilizes a symmetric, U-shaped structure and skip connections to efficiently capture both low and high-level features from input images. Following U-Net, numerous CNN-based segmentation approaches have been presented, e.g., SegNet ~\cite{badrinarayanan2017segnet}, DeepLab~\cite{chen2017deeplab}, PSPNet~\cite{zhao2017pyramid}, and Mask-RCNN~\cite{he2017mask}. Despite the variations in architectural design, the majority of approaches introduced after U-Net expand on its success by either modifying the architecture or offering new performance-enhancing strategies.

The Vision Transformer (ViT)~\cite{dosovitskiy2020image}, as the first successful attempt at employing Transformers in computer vision, has shown potential in several computer vision applications. Yet, there are hurdles with processing high-resolution images due to the high computational cost of processing all image patches. To address this issue, hierarchical structures such as Swin Transformer~\cite{liu2021swin} and Pyramid Vision Transformer (PVT)~\cite{wang2021pyramid} were introduced for the downstream tasks. The shifted-window approach proposed in the Swin Transformer limits self-attention computation to non-overlapping local windows while permitting cross-window attention, thus reducing computational costs.

Following the breakthroughs in Transformer-based architectures for both classification and dense prediction tasks in recent years, a number of studies investigating the capability of Transformers in medical image analysis applications, notably medical image segmentation, have been presented, with TransUnet ~\cite{chen2021transunet}, U-Net Transformer~\cite{petit2021u}, CoTr~\cite{xie2021cotr}, TransFuse~\cite{zhang2021transfuse}, MIXED TRANSFORMER~\cite{wang2022mixed}, and UNETR~\cite{hatamizadeh2022unetr} being among some of the most successful designs introduced on this topic to date. These transformer-based models are built on the self-attention mechanism that allows them to attend to relevant image regions and capture long-range relationships. 

Over the past few years, A number of studies have endeavoured to gain as much advantage as possible from skip connections by employing the attention mechanism~\cite{li2020attention}, redesigning skip pathways in UNet++, introducing Res paths in MultiResUnet networks ~\cite{li2020attention, ibtehaz2020multiresunet}, or incorporating a skip connection module~\cite{wang2022uctransnet} to narrow the semantic and resolution disparities between encoder and decoder representations.

\section{Proposed Solutions}
\label{sec:solutions}
\subsection{Insights into Cervical Spinal Cord and Imaging Modalities}
We propose a multi-parametric method to evaluate the cervical spinal cord with a focus on quantitative measurements in a healthy population to investigate how its microstructural and macrostructural features are correlated~\cite{elahi2022toward}. The MR images used for this analysis are acquired using a 3T scanner across three different MR machines. All centers have used a standardized imaging protocol, providing multimodal MR imaging, including T1-weighted imaging (T1WI), T2-weighted imaging (T2WI), T2*-weighted imaging (T2S), magnetization transfer imaging (MT), and diffusion-weighted imaging (DWI) of the cervical spinal cord. The primary goal is to determine whether there is any significant relationship between macrostructural and microstructural characteristics of the cervical spinal cord in a healthy population. The second objective is to evaluate the influence of gender and the type of MR machines used for acquisition on the correlation between the cervical spinal cord's micro and macrostructural features.

\begin{figure}[ht]
\centering\includegraphics[width=1\columnwidth]{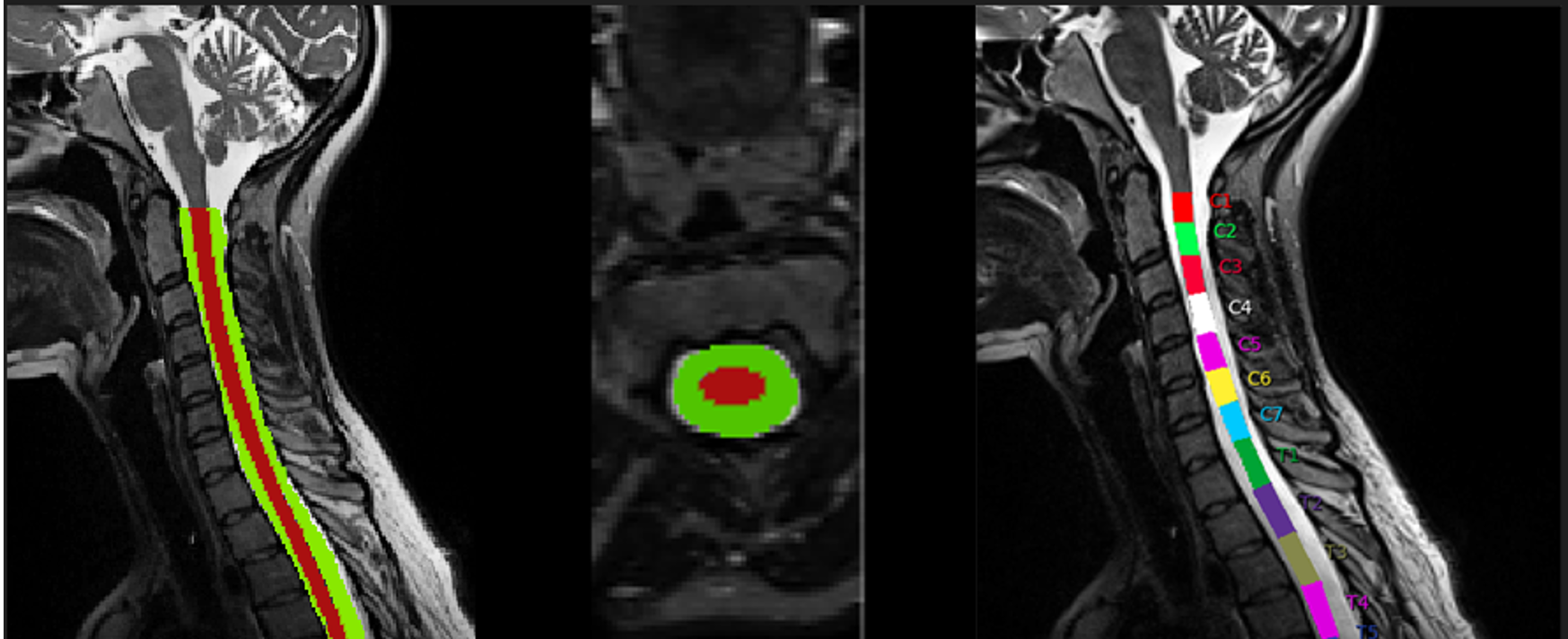}
\caption{Left: cervical spinal cord and cerebrospinal fluid (CSF) segmentation, Middle: cross-sectional view of cord (CSA) and SAC segmentation, Right: per-vertebral level segmentation.}
\label{fig:pipeline}
\end{figure}

The Cervical Spinal Cord Dataset, acquired through the spine-generic protocol, is utilized in this study~\cite{cohen2021generic}. The dataset contains 125 healthy female and 142 healthy male participants from 42 different groups. The imaging data were then fed into the Spinal Cord Toolbox (SCT)~\cite{de2017sct} to estimate the spinal cord cross-section area (SC CSA), the space available for the cord (SAC) and the ratio of SAC/CSA, which is inversely proportional to stenosis at all vertebral levels.
Furthermore, DTI parameters, including the FA, MD, and radial diffusivity (RD), were calculated per-level using the weighted least squares fitting method presented in~\cite{de2017sct}. Figure~\ref{fig:pipeline} represents the processing pipeline, including cervical spinal cord segmentation with a focus on SAC and CSA areas and per-vertebral level segmentation.

We analyzed the per-level extracted metrics to identify any correlation between FA and SAC/CSA, taking into account factors such as the participant's gender and MRI machine type. The results indicate that there is a potential positive Pearson correlation between FA and the degree of stenosis. Moreover, the correlation values vary depending on the type of MRI machine and the gender of participants.
The correlation coefficients are then compared using the z-test on Fisher z-transformed~\cite{hinkle2003applied} correlation coefficients, which allows for a more precise statistical comparison.

\subsection{Deep Learning-based Methodology for Medical Image Segmentation}

U-Net is an extensively adopted and influential medical image segmentation method known for its unique symmetric contracting-expanding architecture, enabling the network to extract both low-level and high-level features from input images. 
As an encoder-decoder architecture, UNet employs a series of convolutional, pooling, and upsampling layers with skip connections, which are designed to concatenate the low-level feature maps extracted from the encoder pathway with their corresponding feature maps in the decoder pathway. This allows the network to retrieve spatial information with a finer granularity that would otherwise be lost due to the downsampling process. 

It has been demonstrated that skip connections are effective in bridging the semantic and resolution gaps between the encoder and decoder features. However, the incompatibility of corresponding representations limits their efficacy. U-Net and other UNet-like variants of encoder-decoder networks used for medical image segmentation suffer from two common challenges: the inadequacy of naive skip connections in modelling long-range correlations between distant pixels, and the structural computational complexity linked to fine-grained prediction tasks.

\begin{figure*}[ht]
\centering\includegraphics[width=0.95\textwidth]{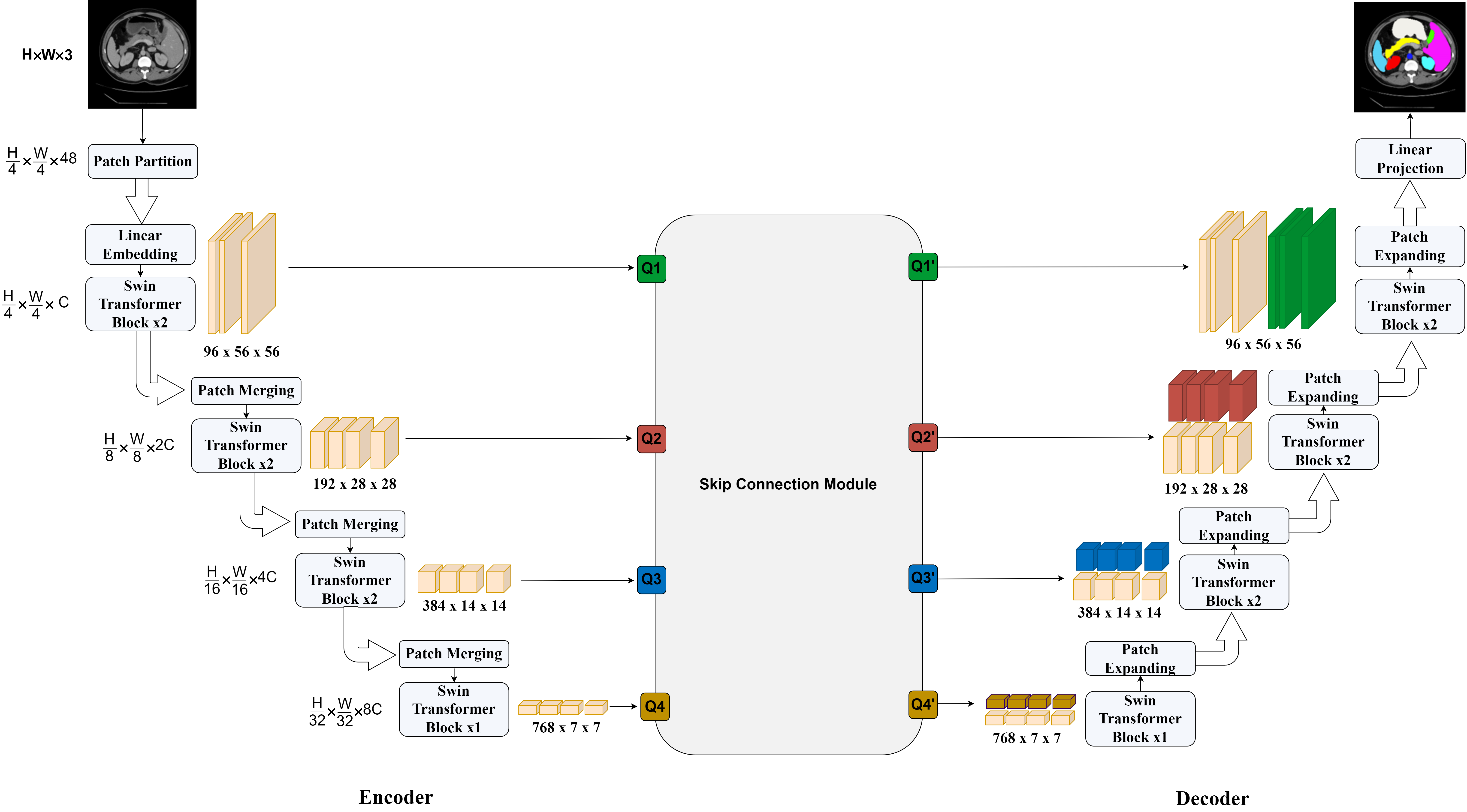}
\caption{The architecture of the proposed SAttisUNet for enhanced medical image segmentation.}
\label{fig:sunet}
\end{figure*}

We propose a deep learning-based medical image segmentation framework called SAttisUNet~\cite{elahi2023sattisunet} to address the aforementioned challenges. As shown in Figure.~\ref{fig:sunet}, we have employed the Swin Transformer as a replacement for convolutional hierarchy in U-Net because of its remarkable aspect of hierarchical architecture and shifted-window technique, benefiting from the strengths of both CNN and Transformer. This design brings greater efficiency while offering the flexibility to model at various scales with linear computational complexity relative to image size. This encoder-decoder network particularly works well when working on dense-prediction tasks involving large variations in the scale of visual entities. 

\begin{figure}[h!t]
\centering\includegraphics[width=0.8\columnwidth]{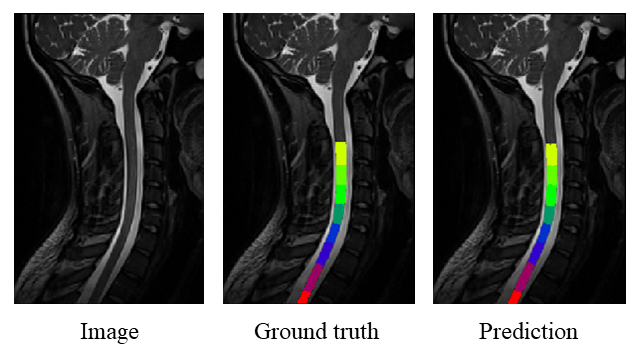}
\caption{The qualitative segmentation results of SAttisUNet on the cervical spinal cord dataset from the sagittal view.}
\label{fig:sunetResult}
\end{figure}

SAttisUNet enriches the encoder’s high-resolution representations prior to integrating them with the decoder’s corresponding feature maps and, therefore, captures fine-grained information more effectively. The features extracted from the Swin Transformer are then fed into the proposed attentive skip connection module, which leverages a merging cross-covariance attention mechanism~\cite{ali2021xcit}. The purpose of this module is to fuse the encoder's features effectively, bridging the semantic gap and ultimately enhancing the overall performance of the segmentation network. We have applied SAttisUNet to the cervical spinal cord dataset with the goal of performing per-level vertebral segmentation. So far, we have achieved an encouraging 89\% segmentation accuracy, given the current scope of data and the stage of experimentation. This result showcases SAttisUNet's adaptability and potential in medical image segmentation tasks. Figure.~\ref{fig:sunetResult} depicts a segmentation result from a sample cervical spinal cord MRI scan.

\section{Expected Contributions}
\label{sec:contributions}
we have surveyed fundamental medical concepts and state-of-the-art segmentation methods (\textit{RQ1}), and analyzed correlations between microstructural
and macrostructural features of the cervical spinal cord (\textit{RQ2}). Moreover, we implemented SAttisUNet, the proposed medical image segmentation method (\textit{RQ3}). 

Future work may include refining and extending the proposed segmentation framework to other medical imaging modalities, such as investigating its efficacy when applied to diffusion-weighted images (DWI). We also plan to investigate the possibility of extending the framework to include classification capabilities. Additionally, refining and exploring modifications to the self-attention mechanism within the encoder-decoder network to better account for both local and global relationships among patches and applying the SAttisUNet model to datasets comprising CSM, thereby expanding its applicability in clinical settings are among the potential improvements and investigations we aim to pursue in our future work. 
Furthermore, we are conducting an ablation study to examine the impact of varying factors, such as the number of skip connections, network depth, and the number of attention heads in multi-head self-attention blocks, on the model's performance. 
Exploring the factors influencing correlations between the microstructural and macrostructural features of the cervical spinal cord may also lead to identifying novel biomarkers for various neurological conditions.

\section{Conclusion}
\label{sec:conclusion}
We have addressed two challenges in the field of medical image analysis: first, objectively analyzing the microstructural and macrostructural characteristics of the cervical spinal cord in a healthy population, and second, enhancing medical image segmentation by overcoming the limitations of naive skip connection strategies, refining the modeling of long-range dependencies, and enriching feature representation. 
Additionally, with segmentation tasks being computationally intensive by nature, we manage the added complexity in our Transformer-based model, thereby improving the macrostructural measurements through the proposed segmentation method. 

In the first part of the research, we conducted an in-depth analysis of cervical spinal cord features in a healthy population. The investigation into the correlation between microstructural and macrostructural characteristics fills a knowledge gap, as limited research has been conducted in this area. The findings contribute to a deeper understanding of the cervical spinal cord's features and highlight the influence of factors such as gender and acquisition machine types on observed correlations. 

In the second part, the proposed UNet-like Transformer-based segmentation framework with attentive skip connections demonstrates improvements compared to conventional UNet-like architectures in medical image segmentation. The integration of the attentive skip connection module and the employment of a merging cross-covariance attention mechanism bridge the gaps between different abstraction levels and capture complex dependencies. Furthermore, incorporating Swin Transformer blocks into U-shaped architectures enhances performance in multi-organ medical image segmentation tasks.

\bibliographystyle{IEEEtran}
\bibliography{references}

\end{document}